\documentstyle{article}
\title{  A class of quasi-stationary regular line
         elements for the Schwarzschild geometry }
\author{ Kayll Lake \\
         Department of Physics \\
         Queen's University \\
         Kingston Ontario Canada K7L 3N6 \\
         lake@astro.queensu.ca }
\date{   3 July 1994 }
\begin{document}
\maketitle
\begin{abstract}
A one-fold infinity of explicit quasi-stationary regular line elements
for the Schwarzschild geometry is obtained directly from the vacuum Einstein
equations. The class includes the familiar Eddington-Finkelstein coordinates,
and the coordinates discussed recently by Kraus and Wilczek.
\end{abstract}

The quasi-stationary spherically symmetric line element in curvature
coordinates (labelled $(t, r, \theta, \phi )$) is given by
\begin{equation}
ds^{2}=-a(r)dt^{2}+2b(r)dtdr+c(r)dr^{2}+r^{2}d\Omega ^{2}
\label{eq:metric}
\end{equation}
where $d\Omega^{2}$ is the metric of a unit sphere
$(d\theta ^{2}+\sin ^{2}(\theta)d\phi ^{2})$.
The term ``curvature coordinates'' is derived from the coefficient of
$d\Omega ^{2}$, and the term ``quasi-stationary'' is derived from the fact
that for the metric (\ref{eq:metric}), $\xi^{\alpha}=\delta ^{\alpha} _{\ t}$
is a Killing vector that is timelike for $a(r)>0$.

The purpose of this note is to solve the vacuum Einstein equations
\begin{equation}
R^{\alpha} _{\ \beta}=0
\label{eq:einstein}
\end{equation}
in terms of the coordinates defining the line element (\ref{eq:metric}). The
system is over determined and any one of $a(r)$, $b(r)$ or $c(r)$ can be
prescribed. We content ourselves here by simply specifying constants.

With $c(r)=c=\mbox{const.}\ge 0$, we solve (\ref{eq:einstein})
and obtain (in geometrical units)
\begin{equation}
ds^{2}=-(1-2m/r)dt^{2}+2\epsilon (1-c+2mc/r)^{\frac{1}{2}}dtdr
    +cdr^{2}+r^{2}d\Omega ^{2}
\label{eq:solution}
\end{equation}
where $\epsilon=\pm 1$. The constant $m$ is the effective gravitational
mass. The constant $c$ has no invariant physical significance. (Whereas the
surfaces $t=\mbox{constant}$ have distinct properties for different
choices of $c$, this is merely a choice of slicing of the spacetime.) With
$0\le c\le 1$, the $r$-$t$ subspaces of the metric (\ref{eq:solution})
are free of coordinate singularities over the range $0<r<\infty$. The
familiar Eddington-Finkelstein coordinates are obtained by setting $c=0$.
The coordinates discussed, and exploited in a modern context, recently by
Kraus and Wilczek \cite{KW} are obtained by setting $c=1$. These coordinates
have an interesting history \cite{history}. I am not aware of a previous
discussion of the class given by (\ref{eq:solution}). (However, in view of the
elementary nature of the problem, and solution, it would be somewhat of a
surprise if the form (\ref{eq:solution}) is actually new!) Whereas with
$0\le c\le 1$ the line element (\ref{eq:solution}) is free of coordinate
singularities, the coordinates are incomplete. The completion can be carried
out for $0\le c\le 1$ exactly analogously to the case $c=0$ that is
thoroughly treated in standard texts.

The procedure that yielded the form (\ref{eq:solution}) fails to give anything
of further interest. In particular, setting $a(r)=1$ (by choice of scale of
$t$) leads, with (\ref{eq:einstein}), only to Minkowski space. The forms
obtained with $b(r)=b=\mbox{const.}\ge 0$ all have the familiar coordinate
singularity at $r=2m$.\\

This work was supported by a grant from the National Sciences and Engineering
Research Council of Canada. The entire calculation reported here was executed
in a few CPU seconds on a notebook computer running GRTensor under MapleV
Release 3. GRTensor runs on any hardware platform which supports Maple and
is available free of charge from the author.

\end{document}